\documentstyle[preprint,aps]{revtex}
\def\ltsim{\mathop{\raise3pt\hbox{$<$}\llap{\lower3pt\hbox{$\sim
$}}}}
\def\gtsim{\mathop{\raise3pt\hbox{$>$}\llap{\lower3pt\hbox{$\sim
$}}}}
\begin{document}
\draft
\preprint{}
\title{Conserving approximation for the three-band Hubbard model:
flat quasiparticle dispersion}
\author{R.\ Putz, R.\ Preuss, and A.\ Muramatsu\cite{alem}}
\address{Institut f\"ur Theoretische Physik,
Universit\"at W\"urzburg, Am Hubland\\
D-97074 W\"urzburg, Federal Republic of Germany}
\author{W.\ Hanke\cite{weha}}
\address{Institute for Theoretical Physics, University of California,
Santa Barbara, California 93106}
\date{\today}
\maketitle
\begin{abstract}
It is shown that the low-energy single-particle excitation-spectrum of 
the three-band Hubbard model at hole-dopings away from half-filling agrees
remarkably well with Quantum Monte Carlo data and spectroscopic experiments
within the framework of a conserving approximation that includes 
self-consistently the interaction with charge, spin, and two-particle 
fluctuations. The dispersion of the quasiparticle-peak obtained from  
the spectral-weight function is flat around the $(\pi,0)$ and $(0,\pi)$
points as has been 
observed in recent angle-resolved photoemission measurement. The significant 
reduction of the quasiparticle-dispersion near the Fermi-energy is due to 
spin-fluctuations, rather than being induced by band effects. 
\end{abstract}
\par
The nature of quasiparticle excitations in the high-$\rm T_c$
superconductors (HTSC) is of central interest in order to characterize the 
normal-state of these materials. Angle-resolved photoemission spectroscopy
(ARPES) \cite{Olson,Mante,Dessau,AbCaGo} shows on the one hand, 
the existence of a Fermi-surface in the normal-state in accordance 
with Luttinger's theorem supporting the conventional Fermi-liquid concept. 
On the other hand, these experiments find a very flat dispersion of the 
quasiparticle excitations near the X=($\pi$,0) and Y=(0,$\pi$) points. 
This extended saddle-point behavior cannot be derived from 
single-particle calculations (Hartree-Fock, LDA \cite{LDA}) which find, in 
contrast, a van Hove bandstructure singularity. The importance of many-body
correlation effects for such a flat quasiparticle dispersion was revealed in 
recent Quantum Monte Carlo (QMC) simulations of both the three-band
\cite{Do2} and the one-band Hubbard model \cite{bulut1}. 
Related interpretations were also advanced recently on the basis of QMC 
simulations of the t-J model \cite{dago}.
However, in spite of the success of the numerical techniques 
in dealing with strongly correlated systems, the physical interpretation of 
purely numerical QMC results is often difficult without reference to more 
transparent theoretical approaches.

In this letter one-particle spectra for realistic parameters \cite{Do2,Do1,Do3}
of the three-band Hubbard model are presented
based on a conserving approximation
\cite{gen2,gen3}. The obtained spectral distribution of the quasiparticle peak 
agrees remarkably well with QMC- and ARPES results. 
The main features of the experimental bandstructure, such as 
the flat dispersion near the Fermi-energy, bandwidth, and Fermi-velocities
are reproduced by our theoretical approach.
Furthermore, in contrast to the QMC simulations, we are able to distinguish 
the influence of different many-body scattering channels.
We can clearly show that magnetic fluctuations are responsible for the drastic 
reduction of characteristic energy-scales in the quasiparticle excitation 
spectrum compared to their mean-field values.
Thus, the flatness has a many-body origin, rather than being induced by
band effects, a result which may have important implications
particularly for mechanisms of superconductivity based on spin
fluctuations \cite{scal,mori,mont1}. The results presented here 
were obtained at much lower temperatures than in 
numerical simulations such that a direct comparison with experiments is 
possible. 

Our analysis is based on 
the two-dimensional three-band Hubbard model \cite{var1,eme}
with the Hamiltonian
$H - \mu N = H_0 - \mu N + H_U$, where
\begin{eqnarray}
H_0 - \mu N &=& \sum_{i,\sigma} (\varepsilon_d-\mu) 
n_{di \sigma}
+ \sum_{j,\sigma} (\varepsilon_p-\mu) 
n_{pj \sigma} \nonumber \\
&+& \sum_{<i,j>,\sigma} t_{ij}
(d^{\dagger}_{i \sigma}p_{j \sigma} + h.c.) \; , \; \mbox{and} \nonumber \\
H_U &=& U_d \sum_i n_{di\uparrow} n_{di\downarrow} \;.
\label{3BH}
\end{eqnarray}
Here $d_{i\sigma }^\dagger$ and $p_{j\sigma }^\dagger$ denote the
electron creation operators for Cu-3$\rm d_{x^2-y^2}$
and O-2p orbitals at sites $i$ and $j$, respectively with spin $\sigma$,
and $n_{di\sigma}=d_{i\sigma }^\dagger d_{i\sigma }$.
The one-electron part $H_0 - \mu N$ is determined by the local orbital
levels $\varepsilon_d$ (Cu) and $\varepsilon_p$ (O) with 
charge-transfer energy $\Delta = \varepsilon_p - \varepsilon_d$,
and a Cu-O hopping $t_{ij}=\pm t$ between nearest-neighbor
Cu-O sites $<i,j>$, where the sign takes into account the phase factors 
for the 2p and 3d$_{x^2-y^2}$ orbitals.
$\mu$ is the chemical potential and the interaction part 
$H_U$ consists of the on-site Coulomb repulsion $U_d$ (Cu). 
The calculations were performed for the parameter set $ U_d=6, \Delta=4$ 
in the hole representation (as everywhere, the energy-unit is $t=1$). 
This choice follows from recent QMC studies \cite{Do2,Do1,Do3}, 
where it was shown that the above parameter set leads to a consistent 
description of several important features of the HTSC. 

Bickers and Scalapino have developed a propagator-renormalized
fluctuation-exchange (FLEX) approximation \cite{gen2} for the one-band 
Hubbard model based on the conserving approximation (CA) of Baym and
Kadanoff \cite{baka} which has resulted in significantly improved
agreement with QMC results \cite{gen3}. Properties of the superconducting 
state in the same model were also studied recently by that method 
\cite{mont2}
The starting-point of this 
self-consistent field solution is the non-interacting Hamiltonian $H_0$ in 
Eq.\ \ref{3BH}. Diagonalization of $H_0$ leads to the eigenvalues 
$E^0_n({\bf k})$ ($n$=1,2,3) which
build up the unperturbed bandstructure. The single-particle
propagator of the non-interacting system $G^0$ can now be expressed in
terms of these energies $E^0_n(\bf{k})$ and the corresponding eigenvectors
$c_{n \nu}({\bf k})$:
\begin{equation}
\Bigl[ G^0({\bf k}, \omega_m) \Bigr]_{\nu,\nu'} = \sum_n \frac{
c_{n \nu}({\bf k}) c^{*}_{n
\nu'}({\bf k})}{i\omega_m-E^0_{n {\bf k}}+\mu}\; ,
\label{green0}
\end{equation}
which becomes a matrix labeled by the orbital indices ($\nu$= d,
$\rm p_x$, $\rm p_y$) of the three-band Hubbard model.
We employ the Matsubara ($T \neq 0$) formalism where the discrete
fermionic frequencies are given by
$\omega_m= (2m+1)\pi T \: ,\;\; m=\{...,-2,-1,0,1,2,...\}$.
The Coulomb interaction $H_U$ is accounted for within the
field-theoretical FLEX approximation by a power series for the
self-energy. This series includes diagrams (symmetric
in particle-hole and particle-particle channels) taking into account the
interaction of electrons with density (d), spin (s), and particle-particle
(p) fluctuations. For the present three-orbital Hubbard model the
single-particle self-energy is as follows:
\begin{equation}
\Bigl[ \Sigma(k) \Bigr]_{\rm dd} = \frac{T}{N} \sum_q 
\biggl\{ \, \Bigl[ G(k-q) \Bigr]_{\rm dd}
\left( V^{(2)}(q) + V^{\rm (d)}(q) +V^{\rm (s)}(q) \right) 
 + \, \Bigl[ G(-k+q) \Bigr]_{\rm dd}  V^{\rm (p)}(q) \ \biggr\} \; ,
\label{self}
\end{equation}
where
\begin{eqnarray}
V^{(2)}(q)  &=& U_d^2 \cdot \chi_{ph}(q) \;, \nonumber\\
V^{\rm (d)}(q) &=& \frac{1}{2} U_d^2 \chi_{ph}(q) 
\left( \frac{1}{1+U_d\chi_{ph}(q)} - 1 \right) \;, \nonumber\\
V^{\rm (s)}(q) &=& \frac{3}{2} U_d^2 \chi_{ph}(q) 
\left( \frac{1}{1-U_d\chi_{ph}(q)} - 1 \right) \;, \nonumber\\
V^{\rm (p)}(q) &=& -U_d^2 \chi_{pp}(q)
\left( \frac{1}{1+U_d\chi_{pp}(q)}-1 \right)\; .
\label{abgesch}
\end{eqnarray}
$k$ stands for the fermionic momenta and Matsubara frequencies 
$({\bf k},\omega_m)$, and $q$ for the bosonic $({\bf
q},\nu_n)$ ones.  $\chi_{ph}$ and $\chi_{pp}$ are the propagators of
non-interacting particle-hole and particle-particle fluctuations:
\begin{eqnarray}
\chi_{ph}(q) & = & \frac{-T}{N}
\sum_{k} \Bigl[ G(k+q) \Bigr]_{\rm dd} \, \Bigl[ G(k) \Bigr]_{\rm dd} 
\;, \nonumber \\
\chi_{pp}(q) & = &  \frac{T}{N}
\sum_{k} \Bigl[ G(k+q) \Bigr]_{\rm dd} \, \Bigl[ G(-k) \Bigr]_{\rm dd} \;.
\label{chiph}
\end{eqnarray}
Inserting this self-energy into Dyson's equation the single-particle
propagator has to be computed self-consistently:
\begin{equation}
\Bigl[ G({\bf k},\omega_m) \Bigr]^{-1} =
\Bigl[ G^0({\bf k},\omega_m) \Bigr]^{-1}-
\Bigl[ \Sigma[G,U]({\bf k},\omega_m) \Bigr]
\; ,
\label{dyson}
\end{equation}
with $G^0$ standing for
the single-particle propagator of the non-interacting system.

The single-particle excitations are determined by the peaks of the
spectral-weight function
\begin{equation}
A({\bf k}, \omega) = - \frac{1}{\pi} \sum_{\nu} \, 
{\rm Im} \Bigl[ G({\bf k}, \omega) \Bigr]_{\nu,\nu} \; .
\end{equation}
For the analytic continuation of the Green's function $G$ to real
frequencies $\omega$ we employ a simple Pad\'{e} algorithm \cite{ViSe}.
All results were obtained on ${\bf k}$-space lattices of 16 $\times$ 16
points in the first Brillouin zone (BZ). 

In Fig.\ \ref{beta10}, we plot the dispersion of the quasiparticle peak
in $A({\bf k}, \omega)$, which is located closest to the chemical
potential $\mu$ along the symmetry lines in the BZ. The hole doping 
($\delta=0.25$ away from half-filling) and the relatively high temperature 
($T=0.1t$) were chosen to compare our FLEX approximation to our best
converged QMC simulations 
which were performed on a 8 $\times$ 8 lattice (192 sites) with periodic 
boundary conditions. The FLEX dispersion for the single-particle excitations 
is in good agreement with the QMC data, in particular, for the peaks in
$A({\bf k}, \omega)$ crossing the chemical potential $\mu$ ($\omega=0$
in Fig.\ \ref{beta10}).

In order to test the influence of the temperature on the spectrum, we
have also performed both FLEX calculations and QMC simulations for the
even higher $T=0.5t$ for various dopings. 
Like in similar investigations for the single-band
Hubbard model at the electronic filling $\langle n \rangle =0.83$ at the
same temperature \cite{BuScaWhi}, we find the $\omega=0$ crossing
in $A({\bf k}, \omega)$ slightly above the flat band region
at $X=(\pi,0)$, whereas this crossing occurs below this region at
$T=0.1t$.
This indicates a strong dependence of the $\omega=0$ crossing on $T$ 
at least in the high-temperature regime. 

For the much lower temperature $T=0.01t \approx 150K$, Fig.\
\ref{akomega} displays the spectral weight $A({\bf k}, \omega)$ for
energies near the chemical potential $\mu$ 
and the three ${\bf k}$-space lines as in Fig.\
\ref{beta10}. The hole doping was set to
$\delta=0.15$ away from half-filling in order to compare with
spectroscopic measurements in the metallic phase of the HTSC.
The pairs of numbers on the right side of each panel have
to be multiplied by $\pi / 8$ to give the corresponding ${\bf
k}$-vectors in the two-dimensional BZ. These curves can be directly
compared with experimental data obtained from angle-resolved
photoemission spectroscopy (ARPES) \cite{Mante,Dessau}. Good
agreement is found for the quasiparticle excitation spectra calculated
by the FLEX approximation and the spectroscopic measurements. The main
features of the experimental data are reproduced by our theoretical
approach: Along the $\Gamma$ --- M high symmetry direction we find a
dispersing quasiparticle peak which crosses through the Fermi-energy
near ${\bf k} = (\pi /2, \pi /2)$. 
The two other symmetry lines in Fig.\ \ref{akomega}
clearly show the existence of the flat band dispersion near X=($\pi$,0),
the most striking result of the ARPES experiments. The quasiparticle
peak of $A({\bf k}, \omega)$ is broadened for excitation energies away
from $\mu$ due to strong correlation effects. In fact, it can be seen that 
not only the position of the maxima in $A({\bf k}, \omega)$ (dispersion)
but also the overall shape of the spectra agrees with the 
experimental results.

The single-particle bandstructure obtained from the
low-lying peak in $A({\bf k}, \omega)$ is depicted in Fig.\ \ref{band15}
in comparison to results of ARPES experiments in Bi2212. 
In order to discern the
influence of different many-body scattering channels, we present the
data of three approximative stages: In the Hartree-Fock (HF) or band-theory
approximation (dashed line in Fig.\ \ref{band15}) 
the HF bandwidth, Fermi-velocities, and the characteristic energy-dispersion 
at ($\pi$,0) are far away from agreement with the experiment.
Charge fluctations given by $V^{\rm (d)}(q)$ in Eqs.\ \ref{self} and
\ref{abgesch} tend to reduce these differences (dotted line in
Fig.\ \ref{band15}), but only the inclusion of both longitudinal and
transverse magnetic fluctuations \cite{frenk}
$V^{\rm (s)}(q)$ leads to a flat quasiparticle band near the Fermi-energy. The
particle-particle channel $V^{\rm (p)}(q)$ contributes only in a minor way to
the full FLEX calculations (full line in Fig.\ \ref{band15}) for the
given parameter set. Therefore, Fig.\ \ref{band15} demonstrates that
spin-fluctuations are responsible for the reduced energy-scales compared to 
mean-field calculations, especially for the flat dispersion near $\mu$ and 
the reduced saddle-point energy at ($\pi$,0). The most dominant contributions 
stem from magnetic modes with a wave-vector ${\bf q}$ near to the 
antiferromagnetic point ($\pi$,$\pi$).
\par
Finally, Fig.\ 4 shows a strong correlation between the energy difference
$\Delta \equiv E(X) - \mu$ and the energy $\omega_{\rm spin}$ 
of the maximum in 
${\rm Im} \chi_{\rm spin} ({\bf q},\omega)$ at the incommensurate wavevector, 
where ${\rm Im} \chi_{\rm spin}$ reaches its largest 
value in the BZ. This suggests 
that for dopings $\delta \ltsim 0.25$, the energy difference $\Delta$ 
is locked-in to the magnetic excitations close to the AF-point \cite{gap}. 
Moreover, we observe in our FLEX-calculations
a linear $\omega$-dependence of ${\rm Im} \; \Sigma$ down to 
approximately the same energy-scale. 
Below this energy-scale, ${\rm Im} \; \Sigma \sim \omega^2$ as expected for a 
Fermi-liquid. Details of these features will be given elsewhere \cite{detail}.

In summary, we presented the single-particle excitation spectrum of the 
three-band Hubbard model within the framework of a conserving approximation.
This many-body approach is based on a self-consistent evaluation of
Dyson's equation including fluctuation-exchange (particle-hole and
particle-particle) self-energy contributions. The results are in remarkable
agreement with Quantum Monte Carlo data and spectroscopic experiments
for low-lying single-particle excitation energies. The low-energy
physics of this FLEX calculation is controlled by many-body fluctuations
in the magnetic scattering channel leading to a significant reduction of
characteristic bandstructure energy-scales from their mean-field values.

We ackowledge helpful discussions with P.\ Dieterich, N.\ E.\ Bickers
and, in particular, D.\ J.\ Scalapino, who also carefully read the
manuscript.
The calculations were performed on the Cray Y-MP EL of the Rechenzentrum
der Universit\"at W\"urzburg and on the Cray Y-MP8 of the
Leibniz-Rechenzentrum M\"unchen and of the
H\"ochstleistungsrechenzentrum J\"ulich. We thank these institutions for their
support. One of us (W.\ H.) acknowledges the hospitality at the ITP in
Santa Barbara and support through the NSF--ITP program.
Furthermore, we gratefully acknowledge support by the Deutsche
Forschungsgemeinschaft project No. Ha 1537/5--2 and by the Bavarian
High--$T_c$ program ``FORSUPRA''.

\begin{figure}[p]
\caption{Quasiparticle dispersion along three cuts through the Brillouin zone
with the
high-symmetry points $\Gamma$=(0,0), X=($\pi$,0), M=($\pi,\pi$).
The FLEX result for hole-doping $\delta$=0.25 and temperature $T=0,1t$
is compared with QMC data.}
\label{beta10}
\end{figure}
\begin{figure}[p]
\caption{Spectral-weight function $A({\bf k},\omega)$ for excitation
energies near the chemical potential and ${\bf k}$-vectors along high-symmetry
lines in the BZ.}
\label{akomega}
\end{figure}
\begin{figure}[p]
\caption{Quasiparticle dispersion along high-symmetry lines in the BZ.
Hartree-Fock (HF) results (dashed line) 
and FLEX calculations which include only the charge scattering channel
(dotted line) and all symmetrically chosen particle-hole and 
particle-particle channels (full line) are compared to ARPES experiments
(open circles: Ref.\ [2], full circles: Ref.\ [3]).}
\label{band15}
\end{figure}
\begin{figure}[p]
\caption{Energy-difference $E(X) - \mu$ and energy of magnetic excitations
$\omega_{\rm spin}$
close to the AF-point as a function of doping.}
\end{figure}

\begin{references}
\bibitem{gap} It should be noted that the temperature $T$=0.01$t
\approx$ 150K is still above the temperature where experimentally the
onset of a spin-gap has been obeserved.
\end{references}
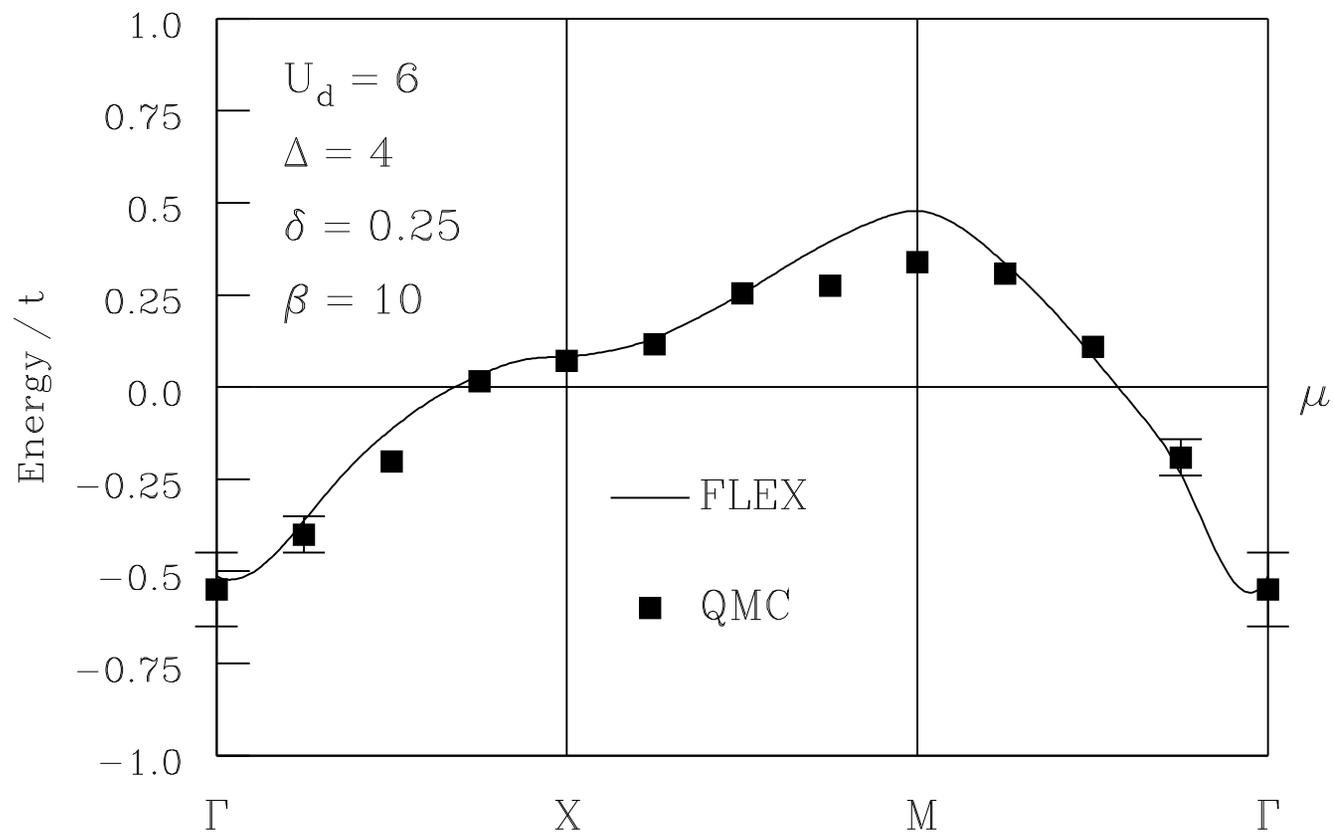
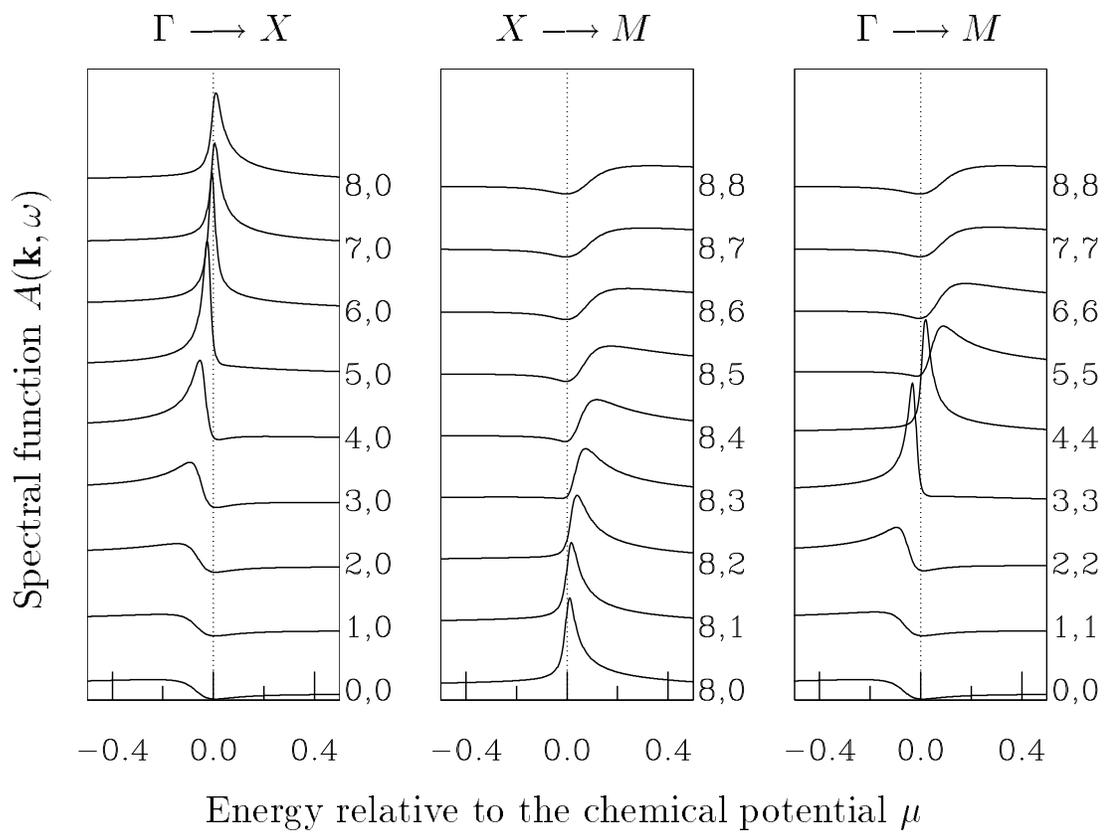
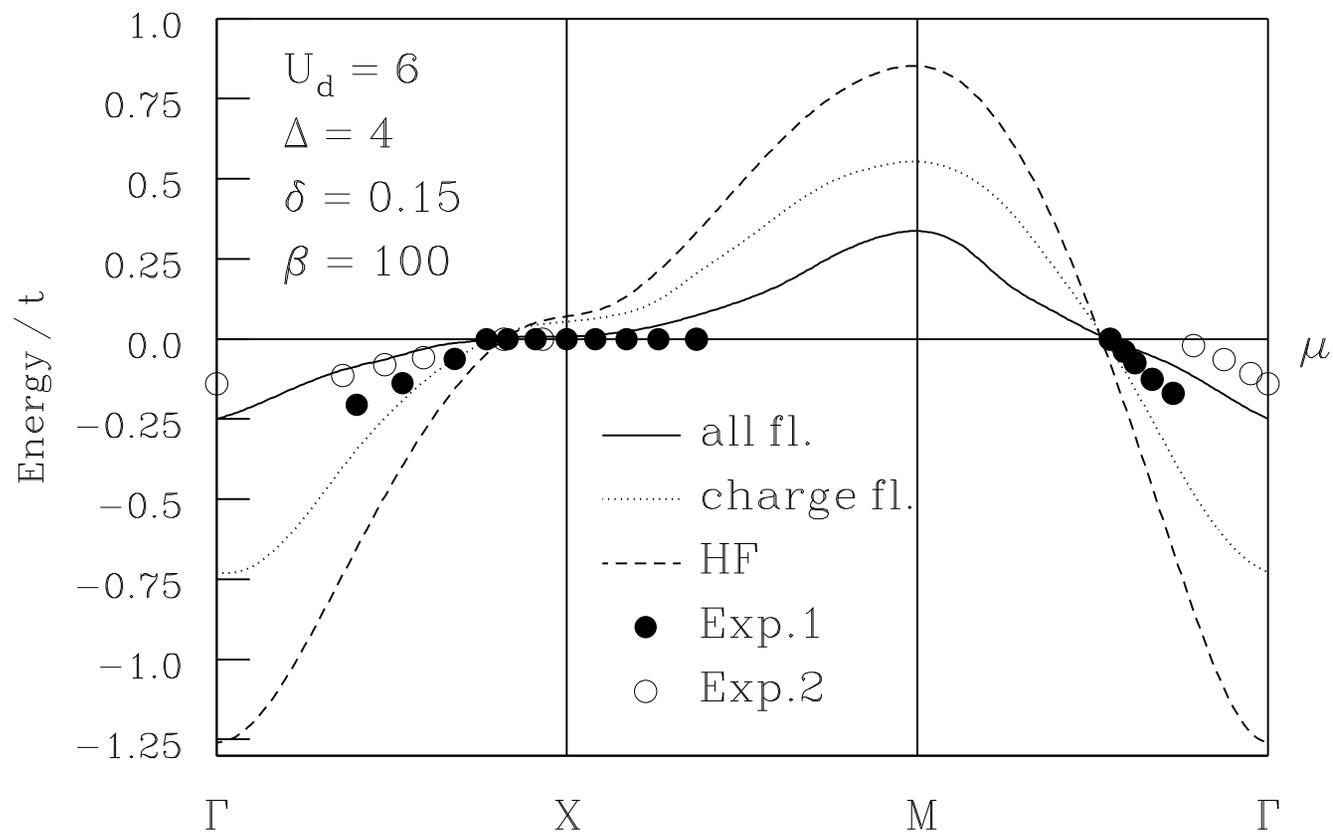
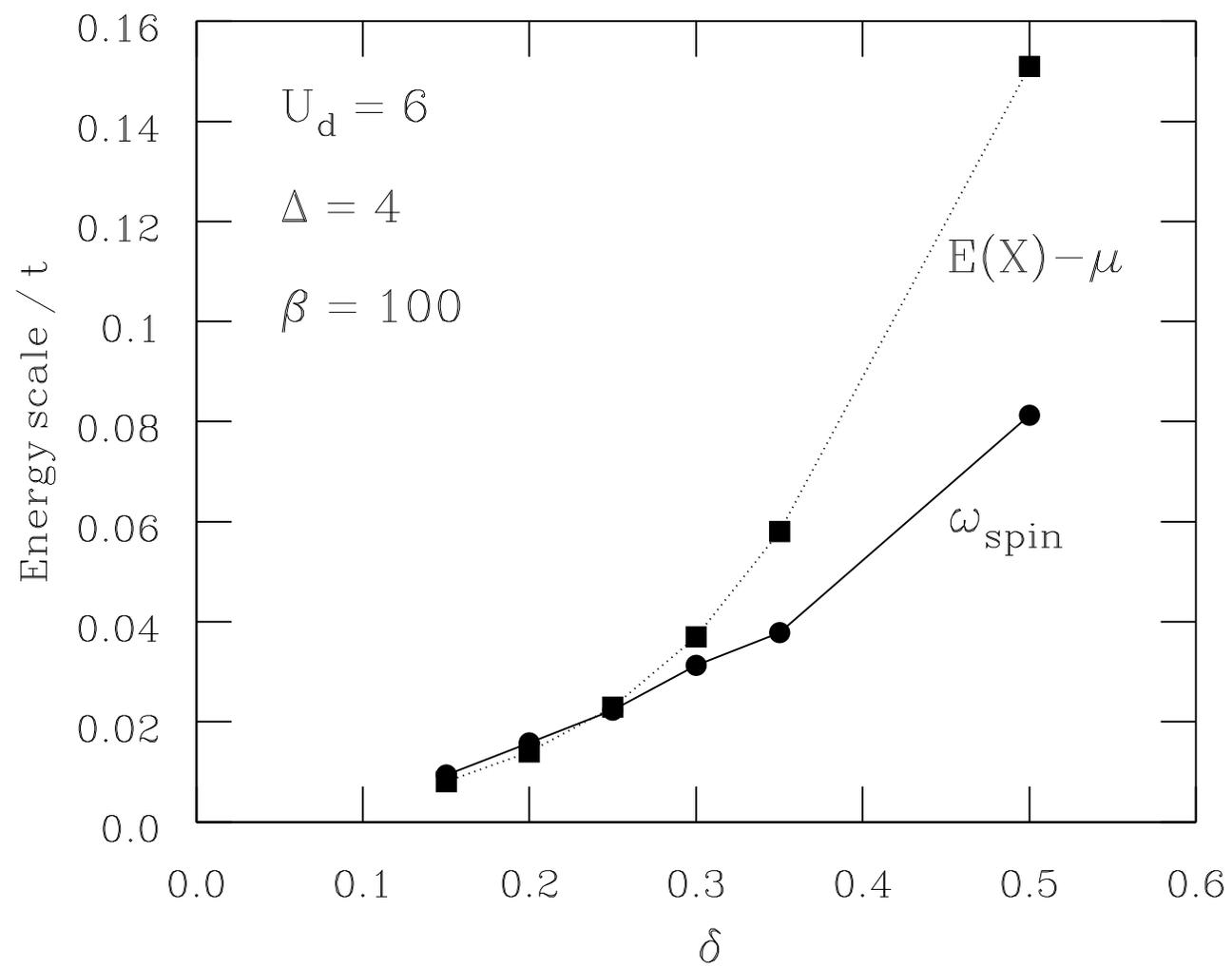
\end{document}